\documentclass[apjl]{emulateapj}
\pdfoutput=1

\submitted{Received 2013 July 9; accepted 2013 July 19}

\shorttitle{Local phenomena at the heliopause}
\shortauthors{Strumik et al.}

\begin{document}


\title{SMALL-SCALE LOCAL PHENOMENA RELATED TO THE MAGNETIC RECONNECTION AND TURBULENCE IN THE PROXIMITY OF THE HELIOPAUSE}


\author{M. Strumik$^1$, A. Czechowski$^1$, S. Grzedzielski$^1$,
W. M. Macek$^{1,2}$, R. Ratkiewicz$^{3,1}$ }
\affil{$^1$Space Research Centre, Polish Academy of Sciences, Bartycka 18A, 00-716 Warsaw, Poland \\
$^2$Faculty of Mathematics and Natural Sciences. Cardinal Stefan Wyszy\'{n}ski University, Dewajtis 5, 01-815 Warsaw, Poland \\
$^3$Institute of Aviation, Al. Krakowska 110/114, 02-256 Warsaw, Poland}




\begin{abstract}
We study processes related to magnetic reconnection and plasma
turbulence occurring in the presence of the heliopause (HP) and the heliospheric
current sheet. It is shown that the interaction of plasmoids initiated
by magnetic reconnection may provide connections between the inner and outer
heliosheath and lead to an exchange of particles between the interstellar medium
and the solar wind plasma shocked at the heliospheric termination shock.
The magnetic reconnection may also cause plasma density
and magnetic field compressions in the proximity of the HP.
We argue that these phenomena could possibly be detected by the
Voyager spacecraft approaching and crossing the HP. These
results could clarify the concepts of the "magnetic highway" and
the "heliosheath depletion region" recently
proposed to explain recent Voyager 1 observations.
\end{abstract}





\section{Introduction}
The interaction of the solar wind (SW) and the local interstellar
medium (LISM) leads to the formation of a cavity in the ambient
interstellar medium called the heliosphere. These two interacting plasma
flows are separated by the heliopause (HP), located between a
termination shock (TS) in the SW and possibly a bow shock (BS) in the
LISM. The region between the TS and HP is called the inner heliosheath
(IHS), whereas the outer heliosheath (OHS) is a region, where a significant
modification of the LISM flow occurs (if the BS exists the OHS is defined as being located
between the HP and BS).

The Voyager 1 and Voyager 2 spacecrafts (V1 and V2 hereafter) crossed the TS in
2004 December at $\sim$94 AU and 2007 August at $\sim$84 AU from the
Sun, respectively \citep{Stoetal05,Buretal05,Stoetal08,Buretal08}.
Both spacecrafts provided magnetic field measurements, but
velocity components of the bulk flow, density and temperature
measurements are available from the plasma experiment for V2 only. For V1
the plasma experiment has not been in operation since 1981, but the flow velocity
components can be estimated indirectly from convective anisotropies of angular
distributions of low-energy ions in the heliosheath
\citep{Decetal05,Decetal07}. Recent measurements by V1 suggest that, at
a distance of $\sim$ 113.5 AU from the Sun the spacecraft entered a
region, where the radial component of the plasma velocity is negative on
average and no mean meridional flow is observed; the region was
identified as a ``transition layer'' \citep{Krietal11,Decetal12}.
Magnetic field measurements from V1 are generally consistent with the
velocity estimations, but the zone was called a ``stagnation region''
as it is part of the IHS rather than the transition layer
related to the HP \citep{BurNes12}.
Very recently, the V1 spacecraft observed two partial depletions in anomalous cosmic ray (ACR)
fluxes, which then decreased to instrumental background at $\sim$ 122 AU from the Sun \citep{Krietal13}.
Hereafter, we will refer to the two partial depletions in ACRs (days 210--214 and 226--233 of 2012)
as the ``precursors'' and the subsequent decrease (day 238 of 2012) as the ``ACR boundary''.
The variations in ACRs were anticorrelated with changes in 
the galactic cosmic rays (GCR) fluxes that significantly increased at this time \citep{WebMcD13,Krietal13,Stoetal13}.
The ACR boundary and its precursors were also associated with sudden increases in the magnetic pressure \citep{Buretal13}.
To interpret the recent V1 measurements, the concepts of the
"magnetic highway" and the "heliosheath depletion region" were proposed to
describe regions of depletion of particles of heliospheric origin
associated with excess of GCRs and magnetic pressure enhancements \citep{Buretal13,Stoetal13}.
In view of the recent V1 observations, detailed understanding of the dynamics of the plasma
and the energetic particles in the vicinity of the HP has become a very timely problem.

One may expect that magnetic topology changes caused by the magnetic
reconnection at the HP could be responsible for providing magnetic
connections between the IHS and OHS \citep{Fahetal86,Swietal10}. In this Letter we test this
hypothesis by numerical simulations of local phenomena in the vicinity
of the HP, additionally including in the model physical features that
are known to likely appear in the IHS: plasma turbulence and possible
occurrence of the heliospheric current sheet (HCS).
Most of the simulations of the reconnection phenomena close to the HP were based on the particle-in-cell (PIC) approach and described
regions of very small spatial size of the order of hundreds of $\lambda_\mathrm{i}=\omega_\mathrm{ci}/V_\mathrm{A} \approx 10^{-5}$ AU,
where $\lambda_\mathrm{i}$ is the ion inertial length \citep{Draetal10,Swietal10,Swietal13}.
We calculate the time evolution of the plasma parameters and the magnetic field
in much larger areas (of linear size 2--20 AU), which implies a magnetohydrodynamical (MHD) approach.
No scaling procedure (similar to that for PIC simulations) is required to interpret our results, but 
obviously our approach provides no insight into small-scale structures on ion inertial length scales.
Our discussion is focused on typical phenomena that could be observed by the Voyager spacecraft.

\section{Model}

\begin{figure}[!htbp]
\begin{center}
\includegraphics[width=8cm]{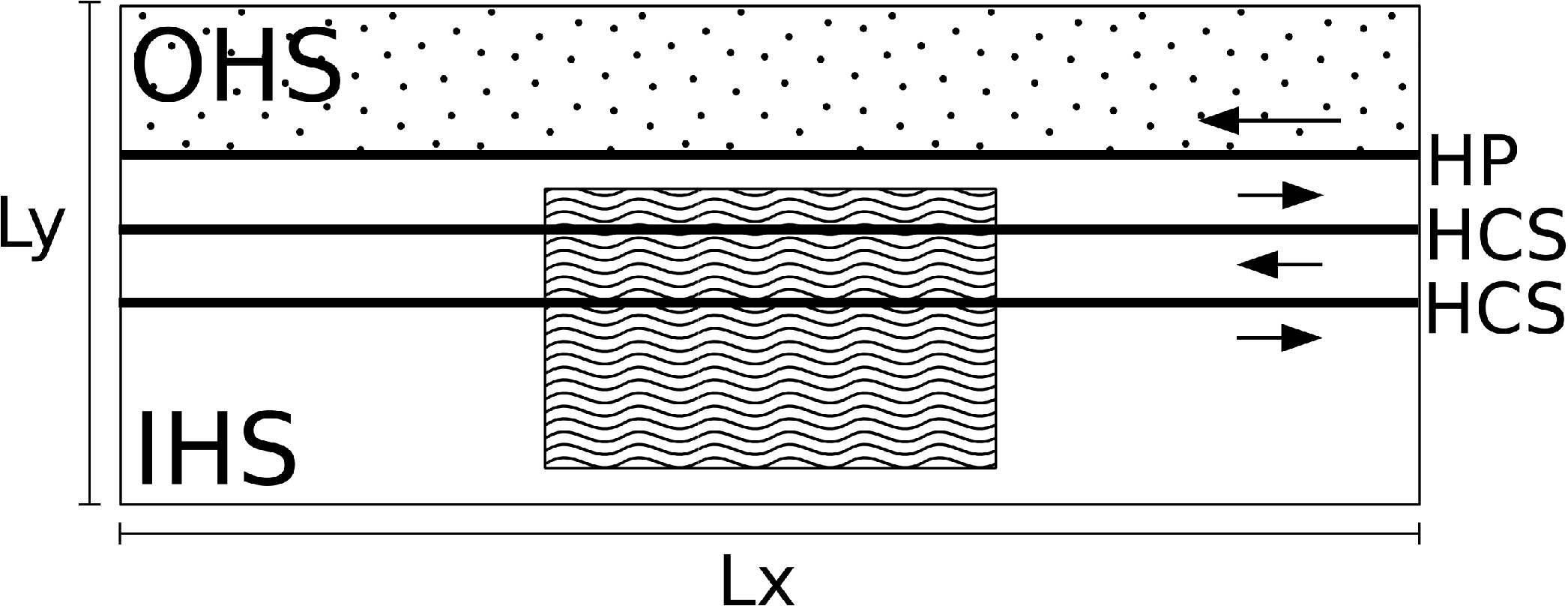}
\caption{
Schematic view of the initial setup for the simulations. The OHS
conditions apply in the dotted region, with the IHS conditions
elsewhere. The thick lines show the positions of the HS and HCS. The orientation
of the magnetic field is shown by the arrows. The region filled by the wavy
pattern is the area of initial injection of randomized velocity
fluctuations or increased noise level.
}
\label{f1}
\end{center}
\end{figure}

The numerical code solves MHD equations in two-dimensional geometry.
Resistive and viscous effects are
not included explicitly but result from small numerical diffusion of the
applied high-resolution MUSCL scheme \citep{KurTad00}. To ensure a
divergence-free magnetic field a flux constrained (staggered mesh)
approach was implemented \citep{BalSpi99}. Neutral particles background
is not included in our model since phenomena related to the interaction
of plasma and neutrals are typically associated with much larger spatial scales.
A schematic view of the initial setup for the simulations is shown in Figure
\ref{f1}.
Open boundary conditions are applied in both the $x$- and $y$-directions.
The resolution of the simulation grid is 5120$\times$512 points,
$L_\mathrm{x}=20$ AU and $L_\mathrm{y}=2$ AU. Density and magnetic field in
the simulation
are normalized to their averages in the IHS,
$N_0$ and $B_0$. Velocity is
normalized to the Alfv\'en speed $V_\mathrm{A}$ in the IHS. Since the
length unit is interpreted as AU, the time unit is $T_0=$AU/$V_\mathrm{A}$.
Estimations discussed below indicate a
rather large value of the Alfv\'en speed $\sim$100 km s$^{-1}$ in the IHS,
which gives the simulation time unit $T_0 \approx$17.5 days.

Using typical recent estimations of plasma parameters in the OHS
(consistently with, e.g. \citet{McCetal12}): plasma density
$N_\mathrm{OHS}=0.05$ cm$^{-3}$, magnetic field strength
$B_\mathrm{OHS}=0.45$ nT, plasma temperature $T_\mathrm{OHS}=6300$ K, we
can compute the ratio $\beta=2 \mu_0 p/B^2$ of the kinetic and magnetic
energy density, which gives $\beta_\mathrm{OHS} \approx 0.1$.
As reported by \citet{Buretal13}, V1 observed a relative increase 
of the magnetic field strength by a factor of $\sim$2 across
the ACR boundary and its precursors. Assuming a
magnetic jump $B_\mathrm{OHS}/B_\mathrm{IHS}=2$ at the HP, the
total (thermal+magnetic) pressure equilibrium condition gives
$\beta_\mathrm{IHS} = 3.4$ used in the initial condition in
our simulations. One should note that $\Delta
\beta=\beta_\mathrm{IHS}-\beta_\mathrm{OHS}=3.3$ is consistent with values
considered by \citet{Swietal10} as resulting from global modeling of the
heliosphere. Recent measurements of plasma density in the IHS by V2 give
$N_\mathrm{IHS}=0.0025$ cm$^{-3}$ \citep{RicWan12}, thus the density
jump $N_\mathrm{OHS}/N_\mathrm{IHS}=20$ was applied at the
simulated HP located at $y$=1.5 AU in the simulation box ($y<$1.5 AU in the
initial condition corresponds to the IHS region).

The IHS presumably contains regions of uniform polarity separated by
regions, where the HCS foldings are closely stacked and magnetic
polarity reversals occur frequently \citep{Czeetal10b}. The most recent
Voyager 1 measurements obtained deep in the IHS show well
separated crossings of the HCS rather than regions of frequently changing polarity
\citep{BurNes12,Buretal13}. As reported by \citet{Buretal13}, the V1 spacecraft
observed a magnetic field reversal (similar to those identified
previously as the HCS crossings) that occurred about 30 days before
the ACR boundary. This time interval corresponds to 
$\sim 0.3$ AU, provided that the observed configuration was time stationary.
To model possible HCS encounters, a Harris-sheet structure is introduced
at $y=1$ AU and $y=1.25$ AU to simulate magnetic field reversals.
Magnetic field reversal is also 
initially arranged at the HP, thus the magnetic reconnection sites are expected
to appear at both the HP and HCS. Antiparallel configuration of the
magnetic field is set up at both the HP and HCS.

Our simulations are of two types. In the first type of simulation
(S1 hereafter) we locally impose small-amplitude random perturbations on
plasma pressure in a confined region of the simulation box. A locally
increased level of noise is supposed to accelerate the growth of the
tearing instability and the development of reconnection sites in a limited
area. In the second type of simulation (S2), a turbulent flow
is initiated by randomized large-amplitude ($\delta V \sim
V_\mathrm{A}$) fluctuations of the velocity vector components. These
fluctuations are injected in the initial condition in a confined region
of the simulation box. The wavenumbers of the velocity perturbations correspond to the largest possible
spatial scales in the simulation (from 0.5 to 2 AU). The dynamics of the system
are then responsible for the development of the magnetic field fluctuations and
the transfer of energy to smaller scales through the turbulent cascade
mechanism. This type of simulation is motivated by observations of the
IHS by the Voyager spacecraft that indicate a significant level of
plasma turbulence \citep{BurNes12}. For both noise- and
large-eddies-initiated simulations of magnetic reconnection the initial
perturbations are applied in the region $8<x<12$ AU and $0.1<y<1.4$ AU and are
initially separated from the simulated HP and boundaries of the
simulation box. Note that the presence of turbulence in the IHS and
the much smaller intensity of fluctuations behind the ACR boundary
is consistent with recent V1 observations \citep{Buretal13}.
The localized character of perturbations considered in our study corresponds
to the local enhancements of turbulence intensity in the IHS. Moreover in
the case of HP, effects related to large-scale curvature of magnetic
field lines can make the shear angle to approach 180$^\circ$, but only in a
confined region, while suppressing reconnection at other locations
(characterized by a smaller shear angle) due to diamagnetic drifts of
the x-line \citep{Swietal10}. Except for the randomized fluctuations
described above, no other mean flow in the simulation box is imposed.

\section{Development of reconnection sites and plasmoids}

\begin{figure}[!htbp]
\begin{center}
\includegraphics[width=8cm]{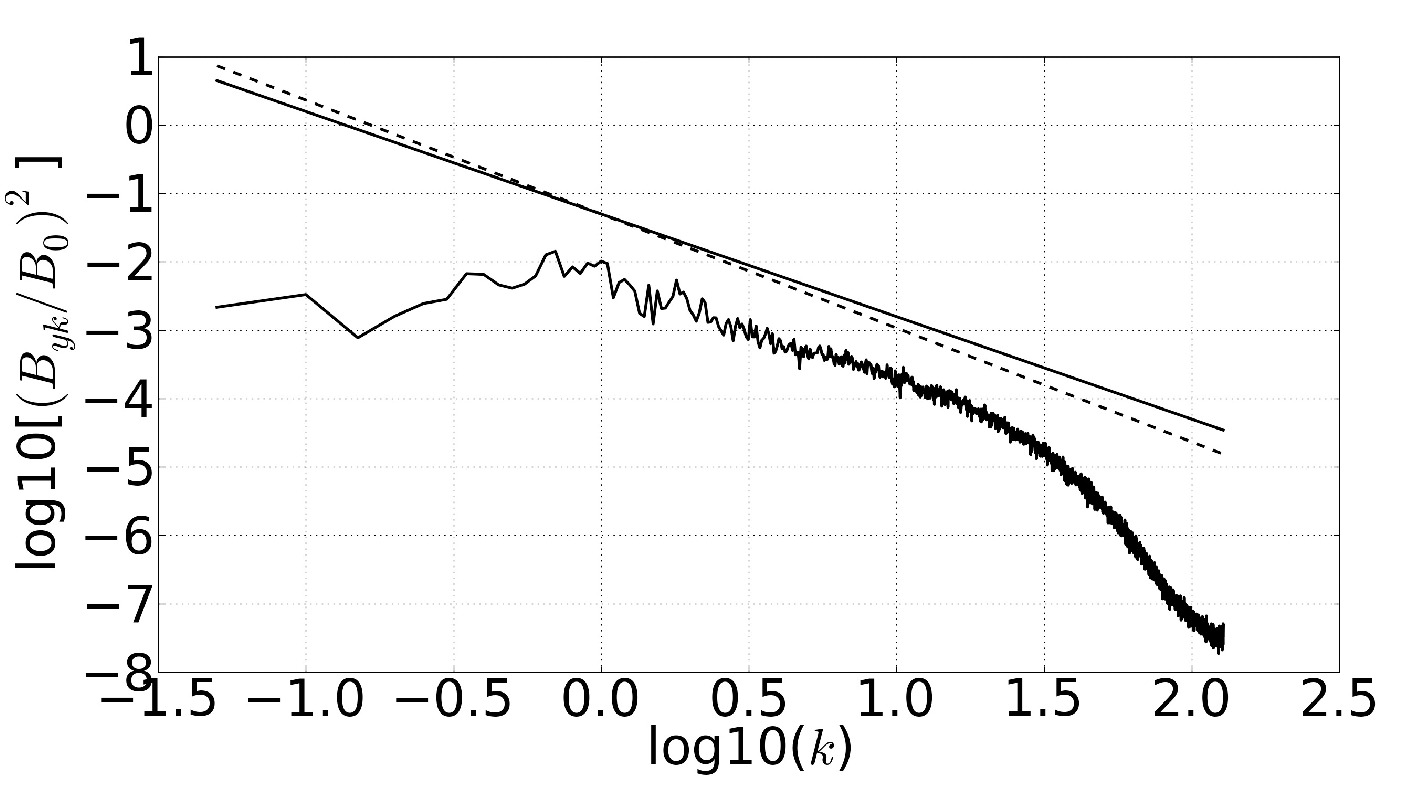}
\caption{Power spectrum of the fluctuations of the magnetic field $B_y$ component for $t/T_0=4$
($B_{yk}$ are Fourier amplitudes corresponding to different wavenumbers $k$).
For comparison, the $x^{-\alpha}$ power law dependence is shown for $\alpha=3/2$ (straight solid line) and
$\alpha=5/3$ (straight dashed line).}
\label{f2}
\end{center}
\end{figure}

Figure \ref{f2} shows the power spectrum of the fluctuations of
the magnetic field $B_y$ component for $t/T_0=4$
averaged over the region of initial injection of large
eddies (S2 simulation case).
One can see that a turbulent cascade develops quickly
and that magnetic fluctuations are excited. The spectrum obtained in the
inertial range is of Iroshnikov--Kraichnan type \citep{Bis03}.

\begin{figure}[!htbp]
\begin{center}
\includegraphics[width=8.5cm]{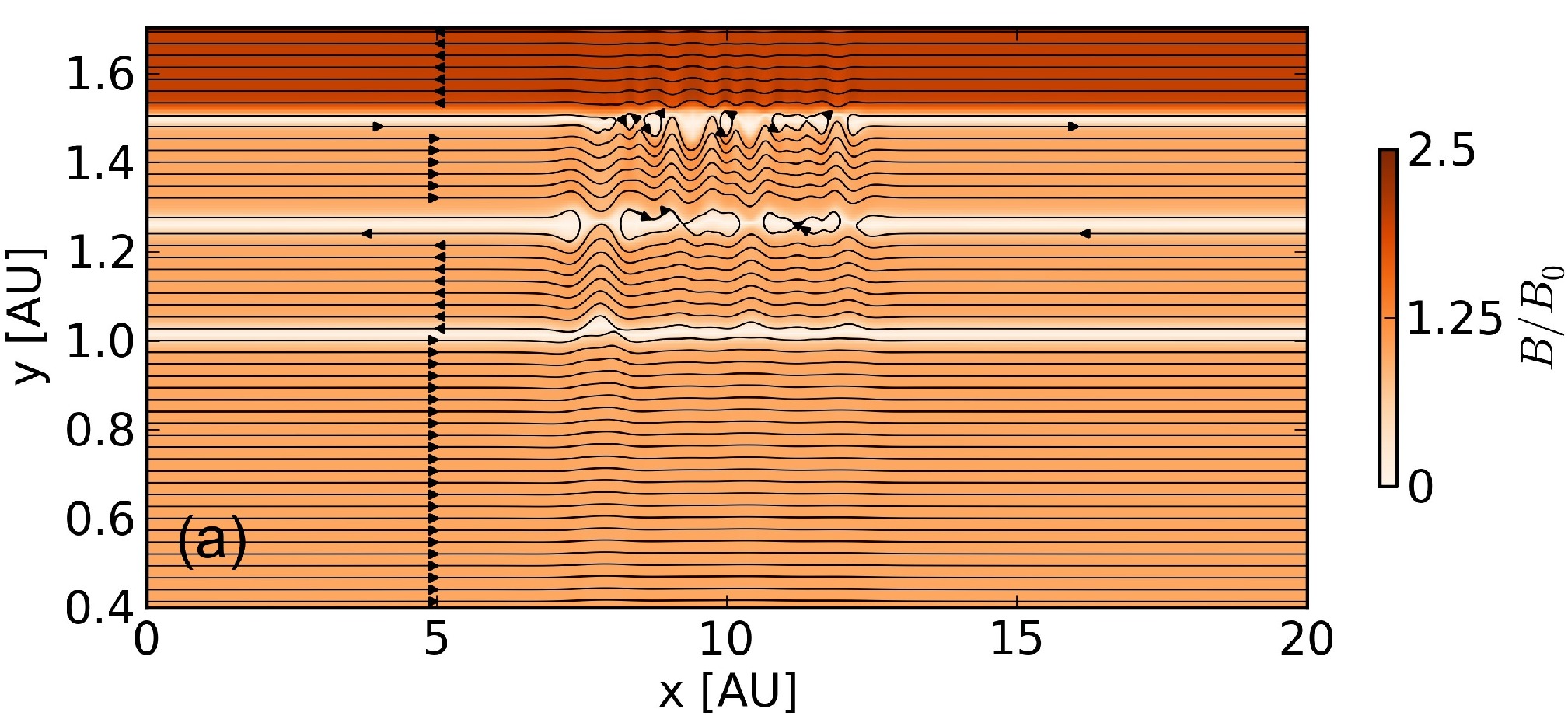}
\includegraphics[width=8.5cm]{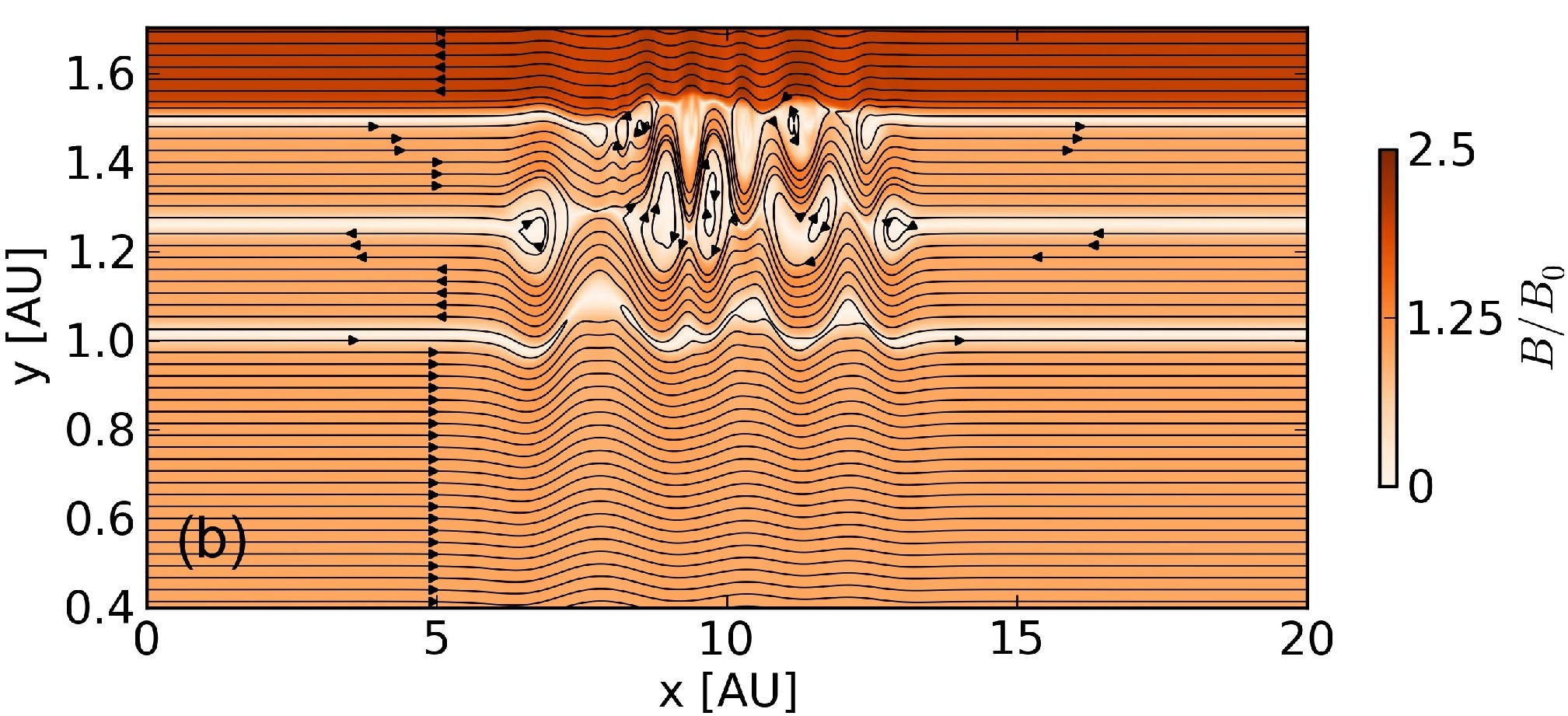}
\caption{Early stages of development of reconnection sites in the
noise-initiated (S1) simulation for (a) $t/T_0=8.5$ and (b) $t/T_0=10$. Magnetic
field lines are superposed on color-coded plots of the magnetic field
strength. Only part of the simulation box is shown and the aspect
ratio of the plot has been altered to better visualize details.}
\label{f3}
\end{center}
\end{figure}

Figure \ref{f3} shows the early stage of the time evolution of the magnetic
field strength and the magnetic field lines in the S1 simulation. For
$t/T_0=8.5$ (see Figure \ref{f3}(a)), reconnection sites appeared due to the
tearing instability mainly at the HP ($y \approx 1.5$ AU) and one of the
HCS ($y \approx 1.25$ AU) located closer to the HP. Topological changes
result in closed magnetic field
lines that indicate the emergence of plasmoids that have just begun to grow.
Later, for $t/T_0=10$ (see Figure \ref{f3}(b)), expanded plasmoids
initiated at different discontinuities are large enough to start to
interact, which results in magnetic compressions in the regions between
interacting plasmoids.
Note that the aspect ratio
of Figure \ref{f3} was altered (elongated in the $y$ direction)
to emphasize the details.
Besides the effects in the neighborhood of the reconnection sites,
the solution at the moments of time illustrated in Figure \ref{f3} remains
laminar, similar to the initial state. Besides the closed magnetic
field lines mentioned above, the reconnection leads to the
appearance of lines that originate and end at the same boundary in the
$x$-direction. In the case of the HP these lines connect the laminar
regions in the IHS with those of the OHS, providing a possibility
of effective transport of higher energy particles between the two regions.

\section{Magnetic connectivity between IHS and OHS}
\begin{figure}[!htbp]
\begin{center}
\includegraphics[width=8cm]{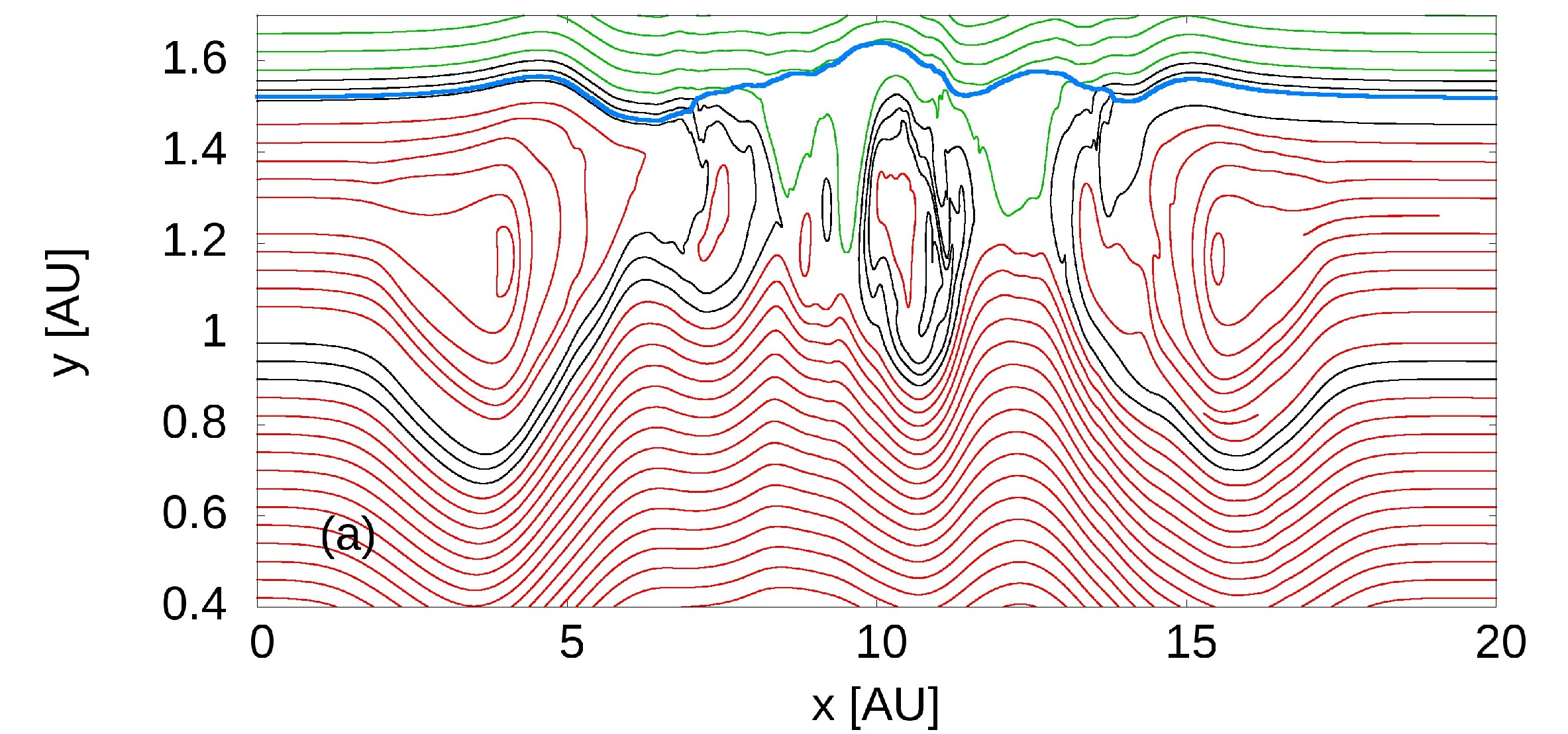}
\includegraphics[width=8cm]{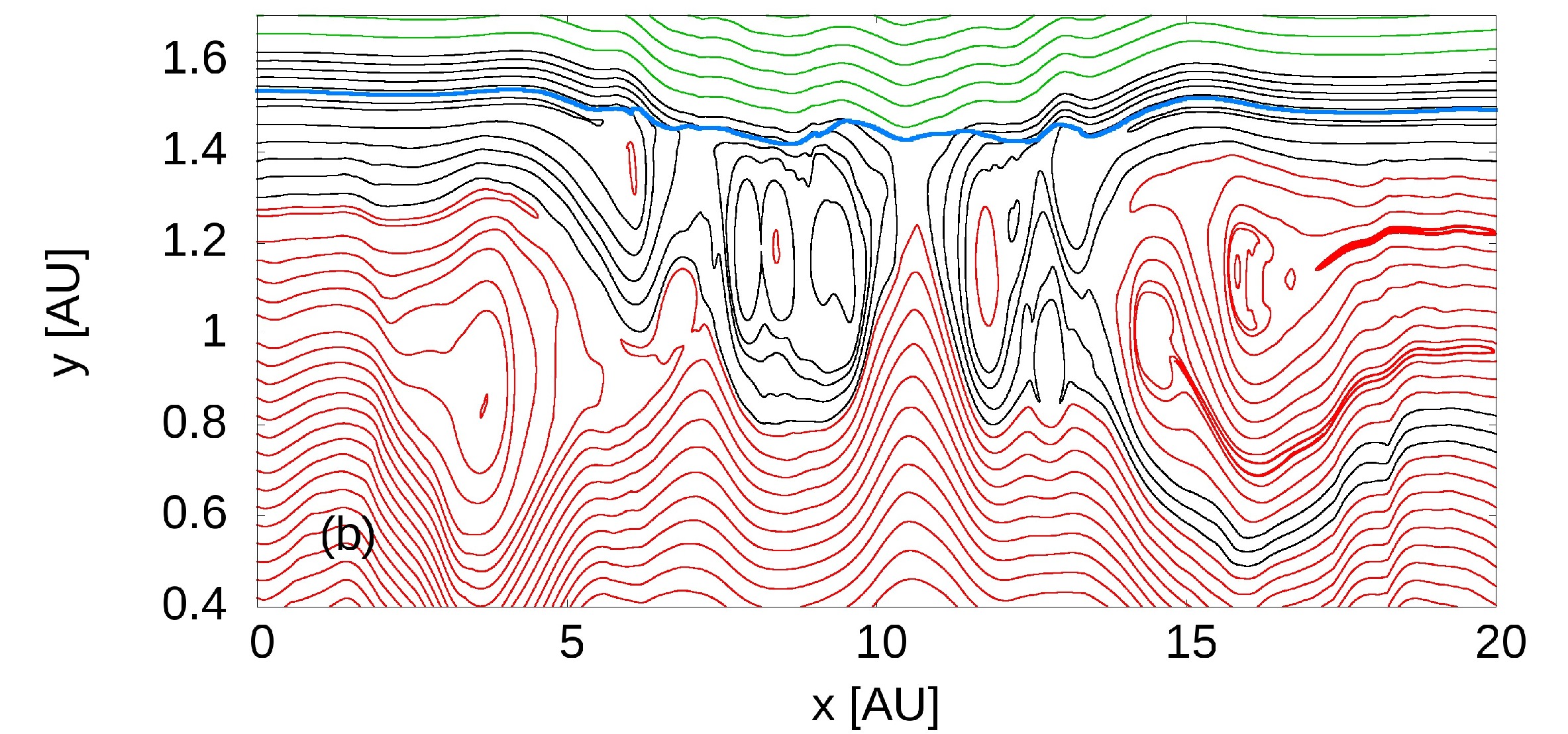}
\caption{Magnetic field lines for (a) S1 simulation (noise-initiated) for $t/T_0=15$ and (b)
S2 simulation (large eddies initiated) for $t/T_0=9$. The different colors show the magnetic connectivity classes:
the green lines are connected to the OHS medium, the red lines to the IHS, and
the black lines provide connections between the IHS and OHS (see the text for
details of the classification procedure).
The blue color marks the line across which a jump in the magnetic field strength occurs.
Only part of the simulation box is shown and the aspect
ratio of the plot has been altered to better visualize details.}
\label{f4}
\end{center}
\end{figure}

The emergence of magnetic connections between
the IHS and OHS is a novel feature revealed
by our simulations that
could be important in the context of recent observations of the V1 spacecraft.
Figure \ref{f4} shows the magnetic field lines classified according to their
connectivity properties.
Different colors are used depending on whether a given line belongs to the IHS,
OHS or connects the two regions.
For the magnetic field lines that originate and end at the boundaries,
the classification is based on the density value at the boundaries.
For closed loops, the plasma density values along the line are used for classification.
The results for the S1 type of simulation for a later moment in time are shown in Figure \ref{f4}(a).
One can see that the laminar regions in the right and left
part of the simulation box shrank significantly in comparison with Figure \ref{f3}.
If we regard the jump in the magnetic field intensity (shown by the blue
line in Figure \ref{f4}) as indicating the HP position in the non-laminar flow,
the black lines in Figure \ref{f4} can be seen to link the IHS with the OHS. The
black lines form bunches that can penetrate quite deeply into the IHS.
Particles streaming along the magnetic field lines are therefore free to
move between the IHS and OHS. A probe moving through the IHS would
therefore likely observe the regions
of increased fluxes of external-origin energetic particles associated with decreased
fluxes of heliospheric-origin energetic particles. These regions would be embedded in normal IHS medium
and could be observed in the IHS quite far from the location of the HP.
Regions
filled by a given class of magnetic field lines shown in Figure \ref{f4}
are often intertwined with regions filled by other classes.
The probe may therefore observe a series of precursors
along its trajectory from the IHS to OHS.
The bunches of black lines in Figure \ref{f4} have typical size of $\sim 0.1$ AU
in the IHS, which is similar to the size of two precursors observed by the V1 spacecraft
\citep{WebMcD13,Buretal13,Krietal13,Stoetal13}.
The results of S2 simulation (large-eddies initiated) shown in Figure
\ref{f4}(b) are qualitatively consistent with the results of the S1
simulation (small-amplitude noise initially) shown in Figure
\ref{f4}(a). This shows the robustness and presumably universal
character of the scenario observed in the simulations, since in the S2 case
reconnection sites have grown in the presence of turbulence, which is
indeed observed in the IHS.

\begin{figure}[!htbp]
\begin{center}
\includegraphics[width=8cm]{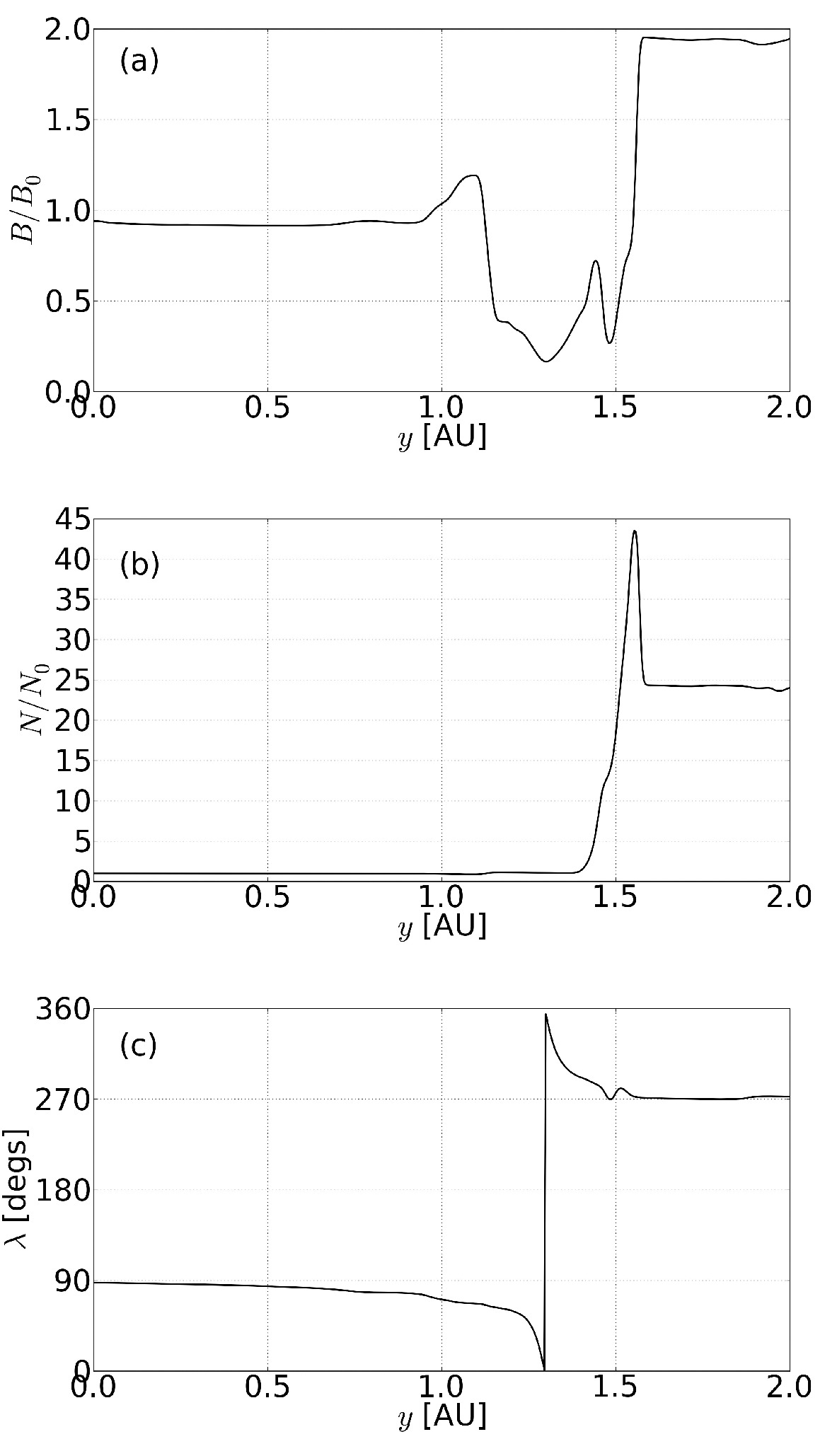}
\caption{Profiles of the magnetic field strength, density, and $\lambda$ angle
along the line $x=7.5$ AU for $t/T_0=17$ in the S1 simulation. The angle $\delta$ is identically zero
for two dimensions.}
\label{f5}
\end{center}
\end{figure}

In Figure \ref{f5} we show profiles of the magnetic field strength,
density and $\lambda$ angle along the line $x=7.5$ AU for $t/T_0=17$ obtained
in S1 simulation. The $\lambda$
angle describes here the orientation of the magnetic field vector with respect
to the $y$-axis of the simulation frame, similar to the $\lambda$ angle
used conventionally in presenting Voyager spacecraft measurements, where
the angle describes deviation of the magnetic field vector from the
radial (R) direction in the radial-tangential (RT) plane of the RTN
frame (see e.g. \cite{BurNes12} and references therein).
The second angle $\delta$ describing the deviation of the magnetic field vector
from the RT plane (or $x$--$y$ plane in calculations) is zero in our simulations due to a purely two-dimensional
configuration assumed in the initial condition.
One can see significant compressions of the density and changes in the magnetic field
strength. The values of the $\lambda$ angle remain in the
proximity of 90$^\circ$ or 270$^\circ$,
except for the region where a single reversal of magnetic polarity occurs.
It is worth noting that only one magnetic
reversal occurs here, whereas three magnetic reversals were set up in
the initial condition for the simulation. This suggests that the
initial configuration of current sheets may change greatly over time,
and the profiles observed by the spacecraft may be misleading when
used to infer about the underlying dynamical processes.
Note that the magnetic field reversal in Figure \ref{f5} is
shifted with respect to the jump in the magnetic field strength, similar to
recent V1 observations \citep{Buretal13}.

\section{Conclusions}

Our simulations clearly show that the concept of the HP as a
well defined boundary separating the IHS and OHS is rather inadequate
when the magnetic reconnection is likely to occur. Magnetic
connections resulting from topological changes related to the magnetic
reconnection may provide the possibility of transporting higher energy
charged particles between the IHS and OHS. On the IHS side the bunches
of magnetic field lines connected to the OHS could be met far ahead of
the jump of the magnetic field strength that naively could be considered
to be an indicator of the HP crossing. The size of the bunches obtained in our
simulations is similar to the size of precursors of the ACR boundary
observed by the V1 spacecraft. We expect that the
simulation results presented in this Letter can be helpful in clarifying
the concepts of the "magnetic highway" and the "heliosheath depletion region"
recently proposed in the context of recent observations of the V1
spacecraft.

The results of our calculations are consistent with PIC simulations
of the magnetic reconnection in the vicinity of the HP, where
scaling arguments were used to link the simulations to V1
observations \citep{Swietal13}.
Our calculations support the interpretation of the precursors and possibly also the ACR boundary
in terms of crossing of a HP-associated non-laminar structure resulting from magnetic reconnection.
Note that possible parallel (or antiparallel)
configuration of the directions of the magnetic field in the IHS and OHS was considered in the
context of interchanging instability causing the flux tubes from the OHS
to penetrate the IHS region \citep{Krietal13}. Also,
some results of the global modeling of the heliosphere suggest that this
quasi-parallel configuration is likely to occur along the V1 trajectory \citep{Swietal10,Swietal13}.


Our simulations indicate that fluctuations related to the turbulent flow in
the IHS region significantly hasten the magnetic reconnection at the HCS
and HP. The simulation starting from small-amplitude noise (S1)
takes 1.7 times longer to reach a developed state in comparison
with the large-eddies-initiated simulation (S2). This effect is
presumably related to local thinning of the HCS caused by turbulent
fluctuations and can be important for the IHS region where turbulence
is observed.

\vspace{-0.5cm}
\acknowledgments
M.S. and W.M. acknowledge support by the Polish National Science Center (N N307 0564 40).
R.R. acknowledges support by the Institute of Aviation, HECOLS project, and ISSI.

\clearpage



\clearpage



\end{document}